\documentclass[allclo]{FBSart}
\usepackage{amsfonts}
\usepackage{amssymb}
\usepackage{epsfig,cite}
\usepackage{wrapfig}

\title{Three-body decays: structure, decay mechanism and fragment
properties}
\author{R.~\'Alvarez-Rodr\'{\i}guez$\:^1$\thanks{\textit{Present
address:} INFN Sezione di Pisa, Largo B. Pontecorvo 3, I-56127 Pisa,
Italy. Electronic address: raquel.alvarez@pi.infn.it.}  ,
A.S.~Jensen$\:^1$, E.~Garrido$\:^2$, D.V.~Fedorov$\:^1$,
H.O.U.~Fynbo$\:^1$, O.S.~Kirsebom$\:^1$} \institute{$^1\:$ Department
of Physics and Astronomy, University of Aarhus, DK-8000 Aarhus C,
Denmark \\ $^2\:$ Instituto de Estructura de la Materia, CSIC, Serrano
123 E-28006 Madrid, Spain}

\runningauthor{R.\,\'Alvarez-Rodr\'{\i}guez, et al.}
\runningtitle{Three-body decays: structure, decay mechanism and 
fragment properties}
\begin{document}

\maketitle
\begin{abstract}
  We discuss the three-body decay mechanisms of many-body
  resonances. R-matrix sequential description is compared with full
  Faddeev computation. The role of the angular momentum and boson
  symmetries is also studied. As an illustration we show the computed
  $\alpha$-particle energy distribution after the decay of
  $^{12}$C($1^+$) resonance at 12.7~MeV.
\end{abstract}

\section{Introduction}

The three-body decay of many-body resonances can be accurately
measured in complete kinematics. Information about the decaying state
and the decay mechanism is usually extracted from the measurement of
the three fragments after the decay. Although this is a common
practice, the situation is ambiguous. The experimental analyses of
these processes are based on the $R$-matrix formalism which inherently
assumes two successive two-body decays. The input are the properties
of the intermediate two-body states and the population of the nuclear
many-body initial state approximated as a three-body system.

Occasionally, in principle contrary descriptions are able to explain
the observed distributions making the understanding of the underlying
physics difficult. An example demonstrating the difficulties is the
$3\alpha$ decay of $1^{+}$ state in $^{12}$C which was successfully
described by two opposite mechanisms: a sequential decay via the
$2^{+}$ state in $^{8}$Be \cite{fyn03}, and a direct decay into the
three-body continuum \cite{alv07}. This requires an explanation. What
information is contained in a full-kinematics measurement of
three-body decay? Apparently unique information can only be extracted
under favorable conditions. The crux of the matter is that the decay
mechanism is related to a ``decay path'', an intermediate structure,
which in contrast to the final state signal in the detector is not an
observable. We shall in this contribution compare the results from
three-body calculations and experimental $R$-matrix analyses.

\section{Energy distributions}

The large-distance observable structure of the many-body initial state
is a three-body continuum state, therefore we compute the resonance
structure in a three-body cluster model \cite{nie01}. We use the
complex scaled hyperspherical adiabatic expansion method to solve the
Faddeev equations which describe the 3-body system. The appropriate
coordinates are the so-called hyperspherical coordinates and consist
of the hyperradius $\rho^2 = 4 \sum_{i=1}^3 (\vec r_i - \vec R)^2 \:$,
and five hyperangles.  In the adiabatic hyperspherical expansion
method the angular part of the Faddeev equations is solved first and
the angular eigenfunctions $\Phi_{nJM}$ are then used as a basis to
expand the total wave function $\Psi^{JM}$.

We include short-range \cite{ali66} and Coulomb potentials. The
many-body effects that are present at short distances are assumed to
be unimportant except for the resonance energy. This is taken into
account by using a structureless 3-body interaction that fits the
position of the resonance.

The resonance wave-function contains information about the decay
mechanism, and the large-distance properties reflect directly the
measurable fragment momentum distributions. The single particle
probability distributions are obtained after integration of the
absolute square of the wave function over the four hyperangles
describing the directions of the momenta.

\begin{wrapfigure}{l}{6.5cm}
\vspace*{-0.5cm}
\begin{tabular}{cc}
\centering \epsfig{file=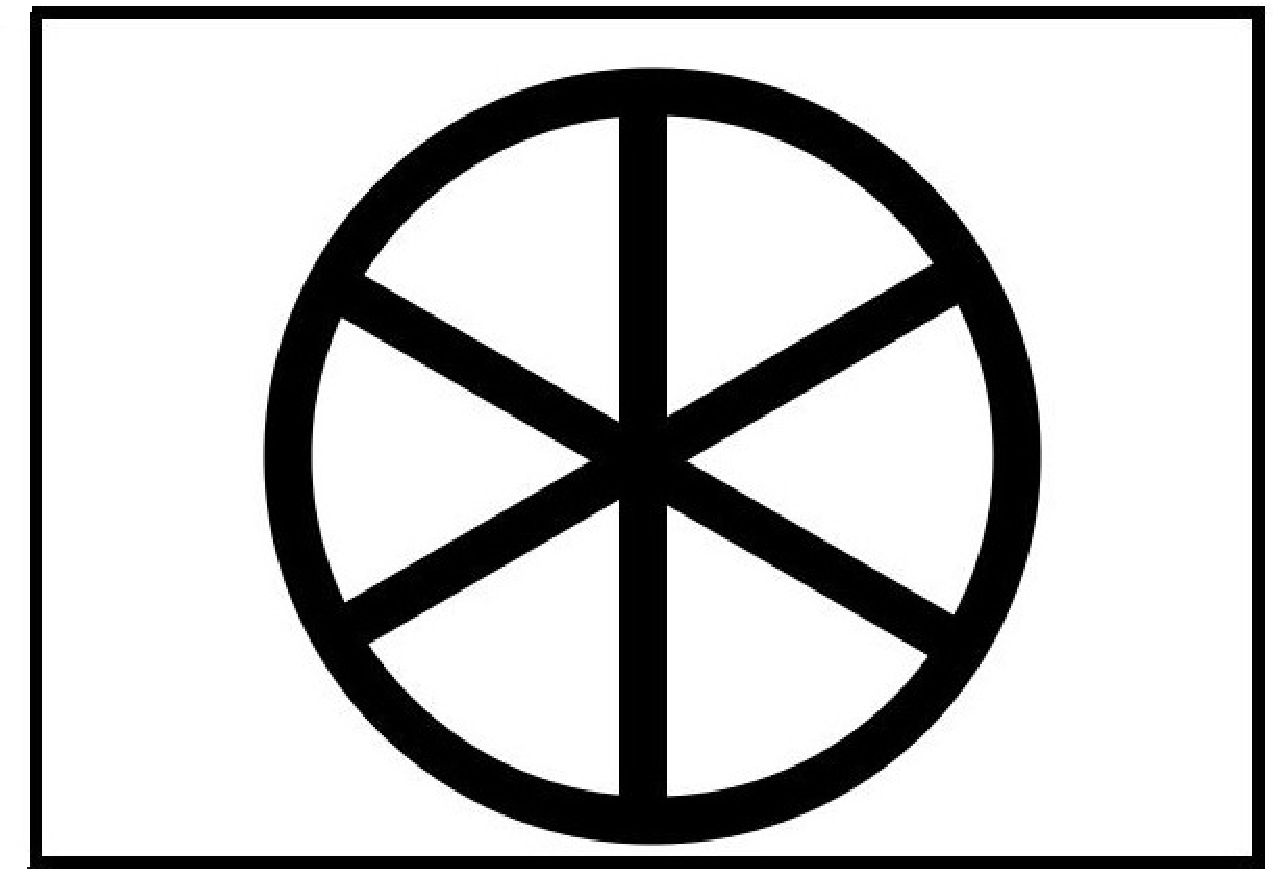,width=0.45\linewidth,clip=}&
\centering \epsfig{file=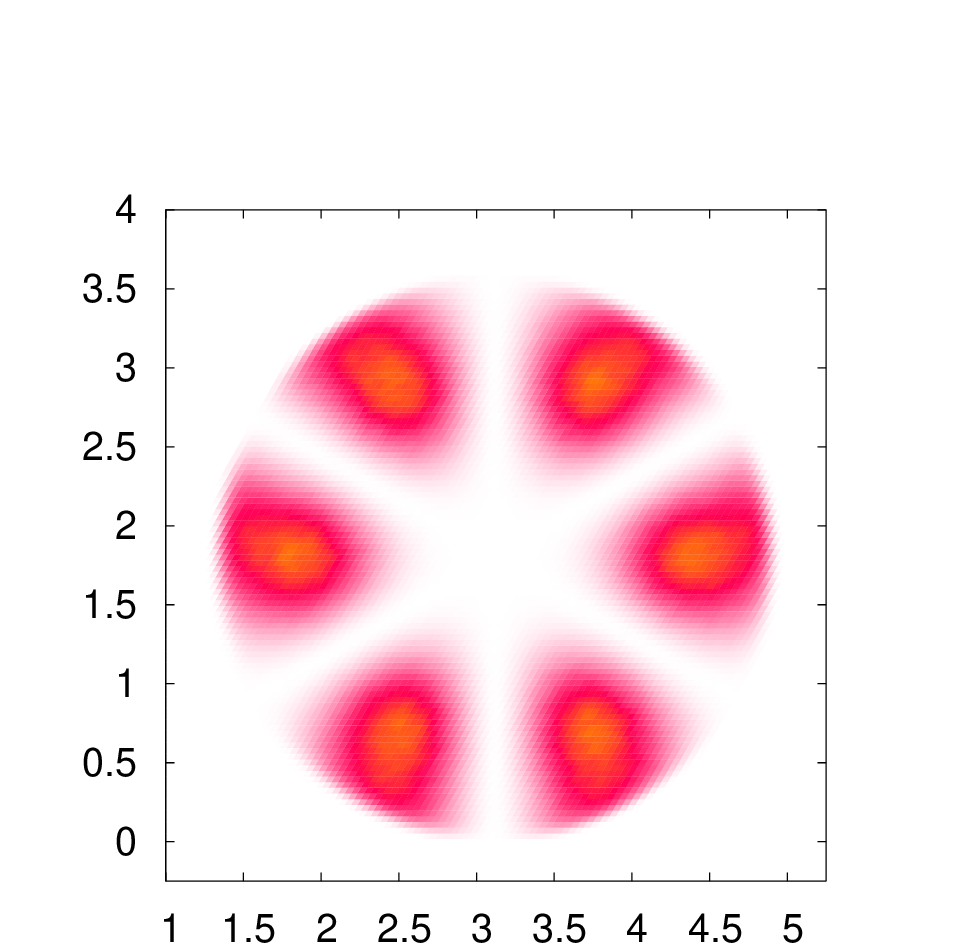,width=0.45\linewidth,clip=}
\end{tabular}
\caption{(Color online) Regions of the $3\alpha$ Dalitz plot where the
density must vanish (in black) for a $1^+$ state (left). Dalitz plot
for the $1^+$ state of $^{12}$C. x-axis corresponds to $(E_{\alpha
1}/2E_{\alpha 2})/\sqrt(3)$ and y-axis to $E_{\alpha 1}$ in MeV.}
\label{sym}
\end{wrapfigure}

The many-body initial state resonance evolves into three clusters at
large distances. The total angular momentum and parity $J^\pi$ is
conserved in the process. This symmetry combined with Bose-statistics
imposes constraints on the resulting momentum distributions. An early
example of these effects applied to three pion decays can be found in
ref.~\cite{zem64}. Fig.~\ref{sym} shows the regions of the Dalitz plot
where the density must vanish for the decay of a $1^+$ state into
three $\alpha$-particles. This gives rise to the minima in the single
$\alpha$-particle energy distribution. The Dalitz plot computed within
the Faddeev framework is also shown and is in agreement with these
symmetry constraints.

\begin{figure}
\centering
\epsfig{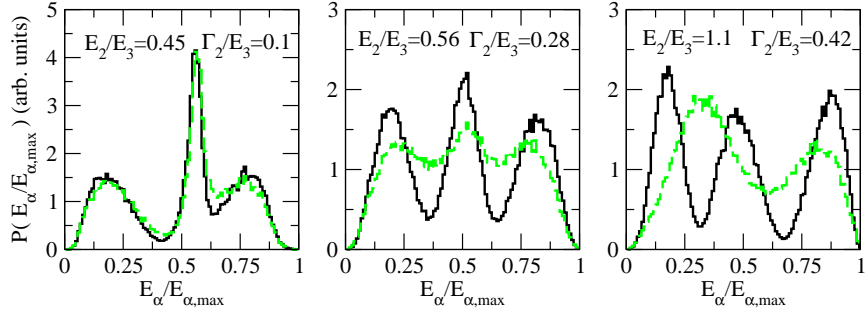}
\caption{(Color online) Single $\alpha$ energy distributions from
R-matrix analysis for the decay into three $\alpha$-particles of the
$^{12}$C($1^+$) resonance at 12.7~MeV of excitation energy. The
energies and widths of the intermediate $^{8}$Be($2^+$) state are
varied as specified in the panels. The green dashed curve corresponds
to the case where the symmetrization of the wave function is
omitted. The middle panel corresponds to the measured resonance
energy.}
\label{rmat}
\end{figure}

Fig.~\ref{rmat} shows the single-$\alpha$ energy distributions,
i.e. the probability for emergence of one $\alpha$ particle with a
given energy divided by its maximum allowed, for the $1^+$ state of
$^{12}$C computed with $R$-matrix analysis \cite{fyn03,fyn}. This
corresponds to the projection of the Dalitz plot in fig.~\ref{sym} on
the y-axis. The decay is assumed to be sequential via $^8$Be($0^+$)
since angular momentum forbids the decay via $^8$Be($2^+$). We have
varied the two-body energy and width. When both the two-body energy
and width are small (left) a narrow peak corresponding to the emission
of the first $\alpha$ arises. The other two $\alpha$'s are related to
the broad peaks. By increasing the two-body energy and width the
three-peak distribution becomes rather pronounced and insensitive to
the two-body parameters when either $E_{2r}/E_{3r}$ is larger than
about 0.5 or the two-body width is large. The same figure contains the
curves corresponding to the case where the boson symmetry is
omitted. For a low and narrow two-body state the effect of this
symmetry seems to be unimportant, but an increase on the width leads
to a two-peak (not three-peak) distribution.

\begin{figure}
\centering
\epsfig{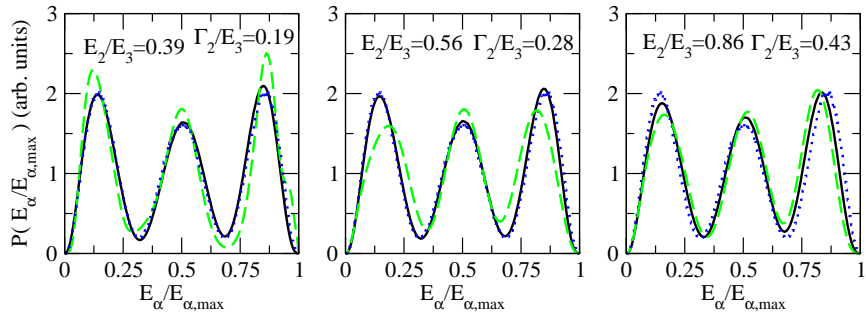}
\caption{(Color online) Single $\alpha$ energy distributions from
Faddeev computation for direct decay of the $^{12}$C($1^+$)
resonance. The three-body energy is varied by changing the strength of
the three-body potential. The relative energies and widths of the
intermediate $^{8}$Be($2^+$) resonance are specified in the
panels. The solid (black) and dashed (green) curves correspond to
complex rotation angles of $\theta = 0.25$ and 0.1 respectively. The
dotted (blue) curve corresponds to democratic decay \cite{kor90}. For
$\theta = 0.25$ only direct decay is shown. }
\label{fad}
\end{figure}

Fig.~\ref{fad} shows the results from the full three-body computation
and the computation from the lowest continuum three-body wave function
(K=8) from ref.~\cite{kor90} (democratic decay). This is the simplest
assumption with the correct symmetries. We have varied the three-body
energy and consequently the three-body width. Two rotation angles have
been considered: one of them is large enough to accumulate the
contribution of sequential decay through the $^8$Be resonance in a
single adiabatic potential, while the other is not. The calculations
include the boson symmetry of the $\alpha$-particles. The results from
the large rotation angle do not include the contribution from the
decay via $^{8}$Be($2^+$) and are very close to the democratic
decay. In the result from the full Faddeev computation the three peaks
are closer to each other and this approaches better the
experiment. The fractions of population at large distance are given in
table~\ref{tab} for the different values of $E_{3r}$ shown in
fig.~\ref{fad}. We can observe that the sequential decay probability
increases as we increase the three-body energy.

\begin{table}
\begin{center}
\caption{The probability $P_{seq}$ for populating the component
related to the decay via $^8$Be($2^+$) at large distances in the
computation of the $1^+$ resonance of $^{12}$C for a complex rotation
angle $\theta=0.25$. The three-body energy ($E_{3r}$) is varied by
adjusting the strength of the three-body potential. The $^8$Be($2^+$)
two-body energy is maintained $E_{2r}=2.7$~MeV. The energies are
referred to the $3\alpha$ or $2\alpha$ separation threshold.}
\begin{tabular}{|l|cccccc|}

\hline \hline $^{12}$C($J^\pi$) & $E_{2r}/E_{3r} $ &
$\Gamma_{2r}/E_{3r}$ & $E_{3r}$ (MeV) & $\Gamma_{3r}$ (MeV) & $l_y$ &
$P_{seq}$\\ \hline & 0.86 & 0.43 & 3.5 & 0.005 & 2 & 0.001 \\ $1^+$ &
0.56 & 0.28 & 5.4 & 0.09 & 2 & 0.12 \\ & 0.39 & 0.19 & 7.8 & 1.15 & 2
& 0.89 \\ \hline \hline
\end{tabular}
\end{center}
\label{tab}
\end{table}
\vspace*{-0.1cm}

\section{Conclusions}

We have computed the observable momentum distributions from decay of
three-body resonances by use of $R$-matrix simulations and from full
Faddeev calculations. We have considered the example of the $1^+$
resonance of $^{12}$C. The angular momentum and boson symmetries
constrain the resulting momentum distributions. The same measured
momentum distributions can be described in different complete basis
sets, e.g. either direct products of two-body states and their
center-of-mass motion relative to the third particle ($R$-matrix) or
three-body continuum wave functions (Faddeev). The fact that different
descriptions seem to work indicates that the same wave function could
be described in different ways. Extracting information of both
structure and decay mechanism can then be misleading and requires
model interpretations. Full Faddeev computations successfully
reproduce the measured distributions.

\end{document}